\documentclass[aps,prl,reprint,amsmath,amssymb,showpacs,nobibnotes]{revtex4-1}
\usepackage{graphicx}
\usepackage{hyperref} 
\usepackage{dsfont}    
\usepackage{color}

\newcommand\trick[1]{}

\begin{document}

\title{Einstein-Podolsky-Rosen  Entanglement of Narrowband Photons from Cold Atoms}

\author{Jong-Chan Lee}
\affiliation{Department of Physics, Pohang University of Science and Technology (POSTECH), Pohang 37673, Korea}

\author{Kwang-Kyoon Park}
\affiliation{Department of Physics, Pohang University of Science and Technology (POSTECH), Pohang 37673, Korea}

\author{Tian-Ming Zhao}
\affiliation{Department of Physics, Pohang University of Science and Technology (POSTECH), Pohang 37673, Korea}

\author{Yoon-Ho Kim}
\affiliation{Department of Physics, Pohang University of Science and Technology (POSTECH), Pohang 37673, Korea}

\date{February 6, 2017}

\begin{abstract}
 Einstein-Podolsky-Rosen (EPR) entanglement introduced in 1935 deals with two particles that are entangled in their positions and momenta.  Here we report the first experimental demonstration of EPR position-momentum entanglement of narrowband photon pairs generated from cold atoms. By using two-photon quantum ghost imaging and ghost interference, we demonstrate explicitly that the narrowband photon pairs violate the separability criterion, confirming EPR entanglement. We further demonstrate continuous variable EPR steering for positions and momenta of the two photons. Our new source of EPR-entangled narrowband photons is expected to play an essential role in spatially-multiplexed quantum information processing, such as, storage of quantum correlated images, quantum interface involving hyper-entangled photons, etc.
\end{abstract}

\maketitle


Entanglement, initially explored experimentally with the polarization states of a pair of photons \cite{Freedman1972,Aspect1981}, has now been demonstrated  in a variety of physical systems, e.g., two spontaneous parametric down-conversion (SPDC) photons \cite{Shih1988,Hong1987}, two-mode squeezed states of optical fields \cite{Ou1992,Ralph1998}, trapped ions \cite{Turchette1998,Riebe2004}, neutral atoms \cite{Hagley1997,Chou2005}, and artificial quantum systems \cite{Bernien2012,Riste2013}. The \textit{gedankenexperiment} proposed by Einstein-Podolsky-Rosen (EPR) in 1935, on the other hand, involves a pair of particles that are entangled in their positions and momenta \cite{Einstein1935,Howell2004,D'Angelo2004}. In addition to fundamental interests, EPR entanglement is essential in quantum imaging and quantum metrology  \cite{Wagner08,Brida10,Perez12,Ono13}. Here we report EPR position-momentum entanglement of narrowband ($\sim$ MHz) photon pairs generated from $\chi^{(3)}$ spontaneous four-wave mixing (SFWM) in a cold atomic ensemble. By using two-photon quantum ghost imaging and interference \cite{Pittman1995,Strekalov1995}, we demonstrate explicitly that the narrowband photon pairs violate the separability criterion, confirming EPR position-momentum entanglement. We further demonstrate continuous variable EPR steering for positions and momenta of the two photons \cite{Reid1988,Reid1989,Mancini2002Entangling,Duan2000,Moreau2014Einstein,Wiseman2007,Cavalcanti2009}. To the best of our knowledge, this is the first experimental demonstration of EPR  entanglement and EPR steering of position-momentum degrees of freedom of narrowband photon pairs, well suited for spatially-multiplexed quantum information processing, storage of quantum images, quantum interface involving hyper-entangled photons, etc \cite{Hale2005,Vudyasetu2008,Boyer2008,Lvovsky2009,Cho12,Nicolas2014}.

The position-momentum-like continuous variable feature of EPR entanglement has been explored initially by using quadrature-phase amplitudes of two-mode squeezed states \cite{Ou1992,Ralph1998}. Genuine EPR position-momentum entanglement of photon pairs became available later by the SPDC process in a bulk crystal  \cite{Howell2004,D'Angelo2004} and is thought to be essential in quantum imaging and quantum metrology \cite{Wagner08,Brida10,Perez12,Ono13}.  The EPR-entangled SPDC photons, however, are inherently broadband, typically on the order of several THz in bandwidth. This large bandwidth makes the source unsuitable for interfacing with quantum memory based on atom-photon coherent interaction, which typically has the working bandwidth of a few MHz.    \cite{Vudyasetu2008,Boyer2008,Lvovsky2009,Cho12,Nicolas2014}.  Although narrowband entangled photon pairs can be generated via cavity-enhanced SPDC \cite{Ou99,Bao08}, the optical cavity necessary for bandwidth narrowing eradicates EPR position-momentum entanglement between photon pairs. Spontaneous four-wave mixing (SFWM) in a cold atom medium can generate narrowband entangled photons without the need for optical cavities \cite{Balic05,Du2008,Cho2014}, but to date no EPR position-momentum entanglement has been reported via SFWM.  In this work, we  demonstrate  EPR position-momentum entanglement of a photon pair generated via cold atom-based SFWM by using quantum ghost interference and ghost imaging. It is shown that the photon pair violates the position-momentum continuous variable separability criterion and satisfies the EPR steering condition  \cite{Mancini2002Entangling,Reid1988,Reid1989,Duan2000}.


\begin{figure*}[t]
\centering
\includegraphics[width=6.8in]{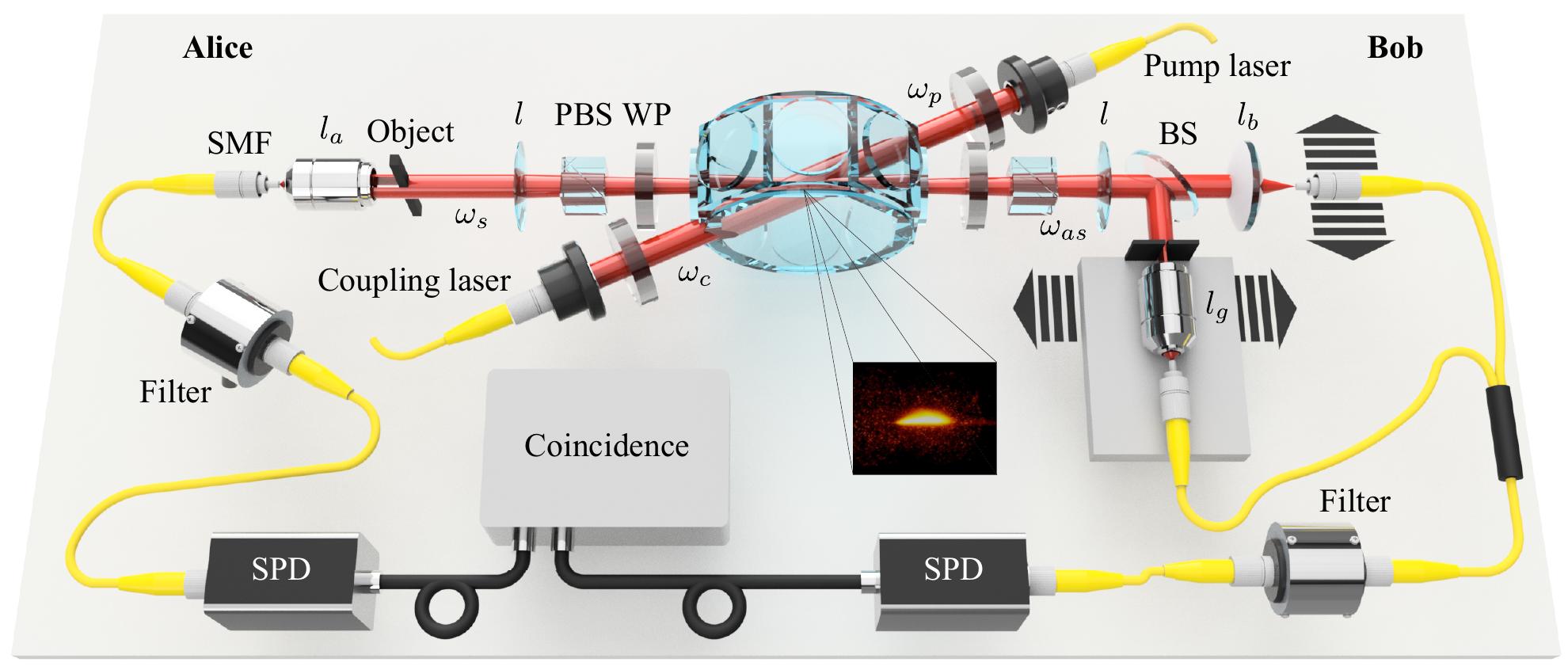}
\caption{\textbf{Schematic of the experiment.} The Stokes ($\omega_s$) and anti-Stokes ($\omega_{as}$) photon pair  with EPR position-momentum entanglement is generated in the 87Rb cold atom cloud by applying the pump ($\omega_p$) and coupling ($\omega_c$) lasers. Waveplates (WP) and polarizing beam splitters (PBS) are used to set the proper polarization states. To optically relay the diverging Stokes and anti-Stokes photons to Alice and Bob, the lenses $l$ are used ($f=400$ mm).  The SFWM photon pairs are measured with single-photon detectors (SPD). SMF and BS refer to the single-mode fiber and the beam splitter, respectively.
}\label{fig1_setup}
\end{figure*}

The experimental schematic is shown in Fig.~\ref{fig1_setup}. The SFWM photon pairs are generated from an ensemble of cold 87Rb atoms in a cigar-shaped 2D magneto-optical trap (MOT) \cite{Du2008,Cho2013,Cho2014}. When the counter-propagating pump ($\omega_{p}$) and coupling ($\omega_{c}$) lasers are applied to the cold atom cloud, the Stokes ($\omega_s$) and anti-Stokes ($\omega_{as}$) photons are  generated via SFWM. The atomic four-level double-$\Lambda$ system used for SFWM consists of $|1\rangle\equiv|5S_{1/2}(F=1)\rangle$, $|2\rangle\equiv|5S_{1/2}(F=2)\rangle$, $|3\rangle\equiv|5P_{1/2}(F=2)\rangle$, and $|4\rangle\equiv|5P_{3/2}(F=2)\rangle$. All the atoms are initially prepared in the ground state $|1\rangle$ \cite{Du2008}. The pump laser is red detuned by $\Delta = 2\pi\times78.5$ MHz from the  $|1\rangle \leftrightarrow |4\rangle$ transition and the coupling laser is resonant to the  $|2\rangle \leftrightarrow |3\rangle$ transition. The SFWM photon pair is collected at the angle of $2.5^{\,\circ}$ with respect to the pump/coupling laser directions and the polarization states of the Stokes and anti-Stokes photons are chosen by wave plates (WP) and polarization beam splitters (PBS). The angle is exaggerated in Fig. \ref{fig1_setup} for clarity. The SFWM photon pairs are measured with single-photon detectors (SPD, Perkin Elmer SPCM-AQRH-13FC) and coincidence events are recorded with time-tagging electronics (SensL HRM-TDC). It is important to note that, to generate a photon pair with EPR position-momentum entanglement, the pump and coupling lasers should not be tightly focused. See Methods for details.

To confirm EPR  entanglement and EPR steering for the position-momentum variables,  we make use of the quantum ghost imaging and interference effects \cite{Pittman1995,Strekalov1995}. Roughly speaking, in ghost interference and ghost imaging experiments with a pair of photons, an object is placed in the path of one photon which is then detected by {a detector with no spatial resolution} and the other photon is  measured  with a scanning detector with spatial resolution. Even though there are no images or interference appearing in the single count rate of the scanning detector, ghost interference or ghost images due to the object occurs in the coincidence count rate of both detectors \cite{Pittman1995,Strekalov1995}.


On Alice's side, we place the object, a metal block of width $=1.23$ mm, in front of the objective lens $l_a$ with focus $f_a=13.5$ mm and numerical aperture (NA) $=0.25$. A single-mode fiber (SMF) is placed at the focus of the objective lens $l_a$ for photon detection. The effective shape of the object, considering the transverse dimensions of the object, the SFWM beam (the Stokes photon), the numerical apertures of the objective lens and the SMF, is a double slit. We thus expect to observe ghost interference and ghost imaging corresponding to the effective double slit placed at the location of the object. The scanning detector is placed on Bob's side. The two-photon ghost interference and ghost imaging measurements require different optical setups for measurement. The transmission or reflection at the beam splitter (BS) selects whether to observe the ghost interference or the ghost imaging, respectively \cite{Howell2004,D'Angelo2004}.
For the ghost interference measurement, the SMF tip is scanned at the focus of the lens $l_b$ with focus $f_b = 25.4$ mm. For  the ghost imaging  measurement, the measurement setup includes a narrow vertical slit of 0.4 mm in width, which defines the imaging resolution, the objective lens $l_g$ ($f_g=13.5$ mm, NA=0.25), and a SMF. The whole setup is mounted on a translation stage and scanned. 



It is well-known that a pair of classically-correlated photons in their positions and in their momenta can lead to ghost imaging and ghost interference, respectively \cite{Bennink2002,Bennink2004,Gatti2004,D'Angelo2005}. It is, however, fundamentally impossible to observe both ghost imaging and ghost interference with a classical position-correlated or momentum-correlated photon pairs \cite{Bennink2004,D'Angelo2005}. On the other hand, if a photon pair is EPR entangled, i.e., quantum correlation exits simultaneously in positions and momenta of the photons, both ghost imaging and ghost interference may be observed  by choosing the appropriate measurement basis \cite{Howell2004,D'Angelo2004}.  Thus, experimental observation of both high visibility quantum ghost interference as well as high contrast quantum ghost imaging from the experimental setup in Fig.~\ref{fig1_setup} can be used to confirm EPR position-momentum entanglement of the photon pair.

\begin{figure}[t!]
\centering
\includegraphics[width=3.4in]{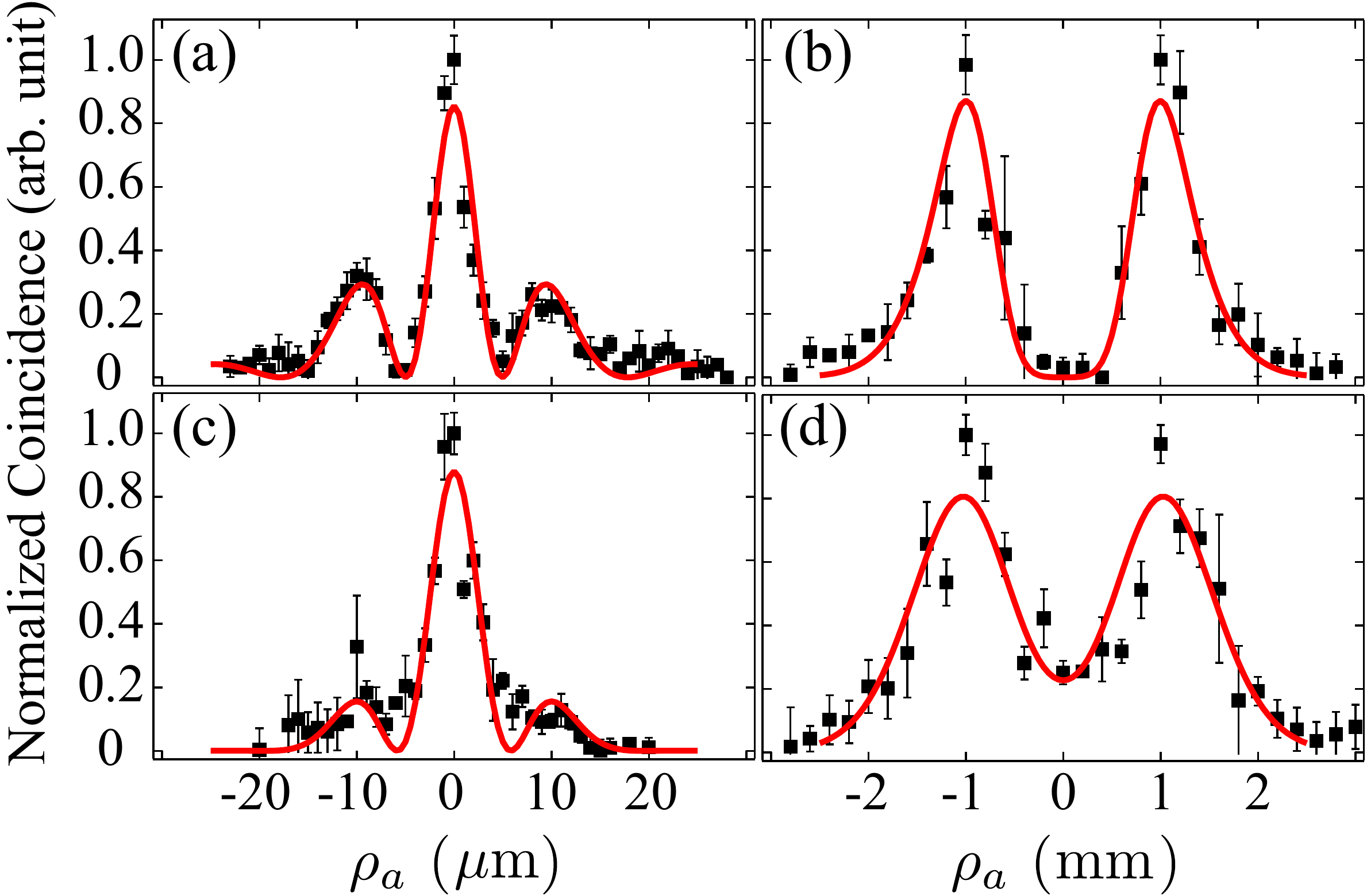}
\caption{\textbf{Experimental results.} When the pump laser is collimated  (beam diameter $2w_0=2.16$ mm), quantum ghost interference (a) and ghost image (b) of the object are clearly observed. From these data, we obtain clear signatures of EPR  entanglement and EPR steering for the position-momentum variables. See text for details. When the pump laser is  focused ($2w_0=235$ $\mu$m), the quality of the ghost interference (c) and ghost image (d) are reduced, signalling reduced EPR entanglement. Each point of the experimental data was accumulated for 60 s.  The solid red lines are numerical fits of the experimental data. Error bars represent statistical error of $\pm$1 standard deviation.}\label{fig2_result}
\end{figure}

The experimental data for quantum ghost interference and ghost imaging are shown in Fig.~\ref{fig2_result}. The coincidence count is normalized to the product of the single counts at the two detectors to remove the effects of single count variations to the coincidence count.  When the pump laser is collimated (beam diameter $2w_0=2.16$ mm), the data clearly exhibit high contrast ghost interference, Fig.~\ref{fig2_result}(a), and ghost imaging, Fig.~\ref{fig2_result}(b), indicating high-degree of EPR position-momentum entanglement. When the pump laser is  focused ($2w_0=235$ $\mu$m), the quality of the ghost interference, Fig.~\ref{fig2_result}(c), and ghost image, Fig.~\ref{fig2_result}(d), are reduced, signaling reduced EPR entanglement. See Methods for details on the exact mathematical shapes of the ghost image and the ghost interference for a given object transfer function. The supplementary information contains more detailed and general calculations for the ghost image and ghost interference. 

To establish EPR position-momentum entanglement between the photon pairs, it is necessary to check if the photon pair violates a separability criterion using the experimental ghost imaging and ghost interference data in Fig.~\ref{fig2_result}. For the transverse positions $(x_1,x_2)$ and transverse momenta $(p_1,p_2)$ of the two particles, if the two particles are in a separable state, they satisfy the inequality \cite{Mancini2002Entangling,Reid1988,Reid1989,Duan2000},
\begin{equation}
\langle(\Delta x_-)^2\rangle\langle(\Delta p_+)^2\rangle\geq |\langle[x_1,p_1]\rangle|^2,
\end{equation}
where $x_-=x_1-x_2$ and $p_+=p_1+p_2$. Experimental violation of the above inequality directly implies that the two photons are in an entangled state. Another notable criterion which we are interested in is the EPR steering inequality \cite{Mancini2002Entangling,Reid1988,Reid1989,Duan2000}. EPR steering is a stricter form of quantum correlation than entanglement such that entanglement is a necessary but not a sufficient condition for EPR steering. Operationally, EPR steering is equivalent to the task of entanglement distribution when one of the two  involved parties is untrusted. Therefore, EPR steering allows for, for example, quantum key distribution when one of the parties cannot trust their device. The EPR steering is possible, or EPR-paradox arises, if the following inequality is satisfied  \cite{Mancini2002Entangling,Reid1988,Reid1989,Duan2000},
\begin{equation}
\langle(\Delta x_-)^2\rangle\langle(\Delta p_+)^2\rangle< \frac{1}{4}|\langle[x_1,p_1]\rangle|^2.
\end{equation}

Our experimental results show strong violation of the inequality in Eq.~(1), hence confirming that the photon pair is EPR position-momentum entangled, and satisfy the EPR steering inequality in Eq.~(2). By fitting the experimental data in Fig. \ref{fig2_result} with the theoretical two-photon correlation functions for ghost interference and ghost imaging, see Methods for details, we obtain the joint uncertainties $\Delta x_-$ and $\Delta p_+$. (Full calculation details are given in the Supplementary Information.) From Fig.~\ref{fig2_result}(a), we obtain $\Delta p_+=1.053\pm 0.635 \hbar$ mm$^{-1}$  and $\Delta x_- = 0.0137 \pm 0.0001$  mm. We thus have $(\Delta x_-)^2(\Delta p_+)^2=0.000208 \pm 0.000177\hbar^2\ll\hbar^2$. Similarly, from Fig. \ref{fig2_result}(b), we have $(\Delta x_-)^2(\Delta p_+)^2=0.000372 \pm 0.000055\hbar^2\ll\hbar^2$. Both results show strong violation of the separability criterion in Eq.~(1) as well as satisfying the EPR steering inequality in Eq.~(2).

To study the effect of spatial profile of the pump to the quality of EPR entanglement of the SFWM photons, we then slightly focused the pump laser  so that the beam diameter $2w_0=$235 $\mu$m at the MOT. The resulting ghost interference and imaging data are shown in Fig.~\ref{fig2_result}(c) and Fig.~\ref{fig2_result}(d), respectively. From Fig.~\ref{fig2_result}(c), we obtain $(\Delta x_-)^2(\Delta p_+)^2=0.0315 \pm 0.0083\hbar^2\ngeq\hbar^2$  and from Fig.~\ref{fig2_result}(d), we obtain $(\Delta x_-)^2(\Delta p_+)^2=0.00326 \pm 0.00124\hbar^2\ngeq\hbar^2$. While both results do violate the separability criterion in Eq.~(1) and satisfy the EPR steering inequality in Eq.~(2), it is clear that the violation of separability in this case is weaker than the previous one in which the pump was collimated.


In summary, we demonstrated, for the first time,  EPR position-momentum entanglement of narrowband photon pairs generated from $\chi^{(3)}$ nonlinearity in a cold atomic ensemble via SFWM. We observed both two-photon ghost interference and ghost imaging effects by using the EPR pair-photon source. From the ghost interference and ghost imaging results, we showed explicitly that the photon pair violates the inseparability criterion as well as satisfying the EPR steering inequality, confirming high-quality EPR position-momentum entanglement between the two narrowband photon pairs. We have also explored the effect of pump spatial profile to the degree of EPR entanglement between the photon pairs. The reported EPR photon pair source is inherently well-suited for efficient interaction and storage in quantum memory/repeater and is expected to play essential role in spatially-multiplexed quantum information processing, including quantum imaging and  quantum metrology.

\section{Methods}

\textbf{Quantum state of SFWM photon pairs.} The two-photon quantum state generated from SFWM can be written as \cite{Du2008,Cho2014}
\begin{eqnarray}
|\Psi\rangle &\propto& \int{d\omega_{as}d\omega_{s}d \vec{\kappa}_{s} d\vec{\kappa}_{as}}\,\chi^{(3)}(\omega_{as},\omega_{s})\, \mathrm{sinc}(\Delta k L/2)\nonumber\\
&\times& \, \mathcal{C}_{\bot}(\vec{\kappa}_+,\vec{\kappa}_-)\,\hat{a}^{\dagger}_{\vec{\kappa}_s} \hat{a}^{\dagger}_{\vec{\kappa}_{as}} |0\rangle,
\end{eqnarray} 
where $\omega_{as}$ and $\omega_{s}$ are the frequencies of anti-Stokes and Stokes photons, $\chi^{(3)}(\omega_{as},\omega_{s})$ is the third-order nonlinear susceptibility of the medium, $\Delta k=(\vec{k}_p+\vec{k}_c-\vec{k}_s-\vec{k}_{as})\cdot \vec{z}$ is the longitudinal phase mismatch along the direction $\vec z$ of the 2D MOT of length $L$, and $\hat{a}^{\dagger}_{\vec{\kappa}_s}$ ($\hat{a}^{\dagger}_{\vec{\kappa}_{as}}$) is the creation operator of photons with the transverse wave vector $\vec{\kappa}_s$ ($\vec{\kappa}_{as}$). Here  $\vec{k}_p$, $\vec{k}_c$, $\vec{k}_s$, $\vec{k}_{as}$ are the wave vectors of pump, coupling, Stokes, anti-Stokes photons within the medium, respectively. The transverse components of the wave vectors are  $\vec{\kappa}_{s}$ and $\vec{\kappa}_{as}$ for the Stokes and the anti-Stokes photons, respectively. The transverse correlation function is $\mathcal{C}_{\bot}(\vec{\kappa}_+,\vec{\kappa}_-)=\tilde{\mathcal{E}}_+(|\vec{\kappa}_+|)\tilde{\mathcal{E}}_-(|\vec{\kappa}_{-}|/2)$, where $\vec{\kappa}_{\pm}=\vec{\kappa}_{as}\pm\vec{\kappa}_{s}$ and $\tilde{\mathcal{E}}_{\pm}$ are envelopes with standard deviations $\sigma_{\pm}$. For a perfectly EPR-entangled photons, $\tilde{\mathcal{E}}_+(|\vec{\kappa}_+|)\rightarrow \delta(\vec{\kappa}_+)$ and $\tilde{\mathcal{E}}_-(|\vec{\kappa}_{-}|)\rightarrow1$ such that the transverse correlation function becomes  $\mathcal{C}_{\bot}(\vec{\kappa}_+,\vec{\kappa}_-)\rightarrow \delta(\vec{\kappa}_+)$. Here, it is assumed that the coupling field is a plane wave with wave vector $\vec{k}_c$ and the medium is larger than the spatial envelope of the pump. Full calculation details are given in the Supplementary Information.

\textbf{Cold atom preparation and SFWM.} The optical depth (OD) of our cigar-shaped MOT was measured to be about 50. The experiment is repeated every 10 ms: 9 ms is used for preparation of the cold atomic ensemble, and 1 ms is dedicated to SFWM for generating narrowband photon pairs. Two different values of pump power are used for two different pumping conditions. When the pump field is nearly collimated with diameter $2w_0=2.16$ mm, the pump power was set at 1.5 mW. When the pump was focused to $2w_0=$235 $\mu$m with a lens of focal length 500 mm, the pump power was 60 $\mu$W. The Rayleigh length in this case was $2z_R=11$ cm, which sufficiently covers the atomic ensemble longitudinally. The coupling field is 3 mW in power and 3 mm in diameter. The polarization states of pump, Stokes, coupling, anti-Stokes fields are chosen to be $\circlearrowright$, $\circlearrowright$, $\circlearrowleft$, $\circlearrowleft$, where $\circlearrowright$ and $\circlearrowleft$ represents right-circular and left-circular polarizations as seen from the receiver, respectively.   To block the  pump and coupling lasers, temperature controlled solid etalon filters (470-MHz full-width-at-half-maximum transmission bandwidth; 21 GHz free spectral range) are placed before the detectors.



\textbf{Ghost interference.} The effective double slit located at the object plane on Alice side causes quantum ghost interference to occur when Bob scans his detector at the far zone, i.e., at the focus of the lens $l_b$. Assuming that in the Alice's measurement plane, her SMF is located at the optical axis defined by the source and the lenses $l$, $l_a$ and $l_b$ ($\vec{\rho}_a=0$), the normalized coincidence count rate $G^{(2)}(\vec{\rho}_b)$ is given by 
\begin{equation}
\left|\displaystyle{\int}{d\vec{\kappa}_{s} \, d\vec{\kappa}_{as}} \, \mathcal{C}_{\bot}(\vec{\kappa}_+,\vec{\kappa}_-)\mathcal{T}(\frac{\lambda_s f}{2\pi}\vec{\kappa}_{s})\mathrm{exp}({-i\frac{f}{f_b}\vec{\kappa}_{as}\cdot\vec{\rho}_b})\right|^2,
\nonumber
\end{equation}
where $\lambda_s$ is the wavelength of the Stokes photons, $f$ is the focal length of the lens $l$, $\mathcal{T}(\vec{\rho}_o)$ is the object transfer function defined by the effective double slit at the object plane $\vec{\rho}_o$. As described in the main text, when the position-momentum correlation is ideal so that $\mathcal{C}_{\bot}(\vec{\kappa}_+,\vec{\kappa}_-)=\delta(\vec{\kappa}_+)$, the two-photon correlation function can be described simply as the square of the Fourier transform of the object transfer function: $G^{(2)}(\vec{\rho}_b)\propto|\widetilde{\mathcal{T}}(\frac{f}{f_b}\vec{\rho}_b)|^2$. 

When the pump has a finite spatial envelope, i.e. $\mathcal{C}_{\bot}(\vec{\kappa}_+,\vec{\kappa}_-)$ is not equal to the delta function, the two-photon correlation function degrades from the ideal Fourier transform.  The shape of the non-ideal ghost interference depends on the two parameters $\sigma_{+}$ and $\sigma_{-}$. By fitting the experimental data to the theoretical calculation, it is possible to obtain $\sigma_{+}$ and $\sigma_{-}$. Using the quantum state in Eq. (3), the uncertainty of the total momentum, $\Delta p_+$, can be calculated to be $\Delta p_+=\hbar \sigma_{+}/\sqrt{2}$. Similarly, the uncertainty of the relative position, $\Delta x_-$, can be calculated to be $\Delta x_-=\sigma_{-}^{-1}/\sqrt{2}$ using the quantum state in Eq. (3). Full calculation details are given in the Supplementary Information.

\textbf{Ghost imaging.} In case of ghost imaging, Bob's detection plane is defined by a narrow vertical slit and ghost imaging is obtained by horizontally scanning the whole measurement setup mounted on a translation stage. Assuming that the opening of Bob's narrow slit is located at $\vec{\rho}_b$ and the Alice's SMF is located at $\vec{\rho}_a=0$, the two-photon correlation function for the ghost imaging set-up can be written as,
\begin{equation}
\resizebox{.48 \textwidth}{!}
{
$G^{(2)}(\vec{\rho}_b)\propto
\left|\displaystyle{\int}d\vec{\kappa}_{s}d\vec{\kappa}_{as}
\mathcal{C}_{\bot}(\vec{\kappa}_+,\vec{\kappa}_-)\mathcal{T}(\frac{\lambda_s f}{2\pi}\vec{\kappa}_{s})\delta(\vec{\kappa}_{as}-\frac{\omega}{cf}\vec{\rho}_b)\right|^2.$
}
\nonumber
\end{equation}
Again, when the pump field is a plane wave so that the position-momentum correlation is perfect, i.e. $\mathcal{C}_{\bot}(\vec{\kappa}_+,\vec{\kappa}_-)=\delta(\vec{\kappa}_{+})$, the two-photon correlation function is reduced to, $G^{(2)}(\vec{\rho}_b)\propto\left|\mathcal{T}(-\vec{\rho}_b)\right|^2$, which is proportional to the square of the object transfer function itself, $|\mathcal{T}(\vec{\rho}_o)|^2$. When the momentum of the pump is not a delta function, the two-photon correlation function has to be calculated from Eq. (3), which can give somewhat blurred image of the object. Again, by fitting the experimental data to theory, we can obtain $\sigma_{+}$ and $\sigma_{-}$. Full calculation details are given in the Supplementary Information.


\section*{Acknowledgements}

This work was supported by Samsung Science \& Technology Foundation under Project Number SSTF-BA1402-07. 

\end{document}


\title{Supplementary Information \\  ``Einstein-Podolsky-Rosen  Entanglement of Narrowband Photons from Cold Atoms''}
\author{Jong-Chan Lee}
\email{spiritljc@gmail.com}
\affiliation{Department of Physics, Pohang University of Science and Technology (POSTECH), Pohang 37673, Korea}

\author{Kwang-Kyoon Park}
\affiliation{Department of Physics, Pohang University of Science and Technology (POSTECH), Pohang 37673, Korea}

\author{Tian-Ming Zhao}
\affiliation{Department of Physics, Pohang University of Science and Technology (POSTECH), Pohang 37673, Korea}

\author{Yoon-Ho Kim}
\email{yoonho72@gmail.com}
\affiliation{Department of Physics, Pohang University of Science and Technology (POSTECH), Pohang 37673, Korea}

\date{February 6, 2017}

\begin{abstract}
Here, we describe in detail the theoretical calculation of two-photon correlation functions for ghost interference and ghost imaging as described in Fig. 1 of ``Einstein-Podolsky-Rosen  Entanglement of Narrowband Photons from Cold Atoms''. 
\end{abstract}

\maketitle

\section{Position and momentum of a photon}

Before we begin, it is required to carefully define the position and momentum we use in the paper, as photons in general does not follow Schr\"{o}dinger's equation as typical free particles with mass. Here, we define the position and momentum variables in the plane \textit{transverse} to the direction of light propagation. The Hilbert space that describes the transverse spatial degree of freedom of photons is isomorphic to the Hilbert space that represents the quantum state of a point particle in two dimensions \cite{Lvovsky2009,Tasca2011,Lee2012}. Therefore, in the transverse space, the position and momentum variables can be defined in analogy to those of a free particle with mass. Also, here we use scalar transverse position and momentum without loss of generality, as the two orthogonal transverse components inherently have no difference.

In classical electromagnetism, the momentum of the electromagnetic field is defined as the volume integral of the energy flux density divided by $c^2$. The wave vector of the field has the direction of the energy flux, and the magnitude proportional to the frequency $|\vec{k}|=\omega/c$. The momentum of a photon is then defined as $\vec{p}=\hbar\vec{k}$, which is a well-known quantum mechanics relation. In the following discussion, we will often use $\vec{k}$ as the momentum of a photon to simplify the notation, which does not affect the result of the measurement.

\section{Second-order correlation function}

The properties of a system of photon pair can be studied by the coincidence of the detection events of two single-photon detectors.
The coincidence count rate is the well-known Glauber's formula for the second-order correlation. For a quantum system described by a density matrix $\vec{\rho}_{12}$, the joint probability of detecting two photons at two space-time coordinates, $(\vec{r}_{1},t_{1})$ and $(\vec{r}_{2},t_{2})$, can be written as \cite{Glauber1963},
\begin{widetext}
\begin{eqnarray}
G^{(2)}(\vec{r}_1,\vec{r}_2,t_1,t_2)=\mathrm{tr}[E^{(-)}_1(\vec{r}_1,t_1)E^{(-)}_2(\vec{r}_2,t_2)E^{(+)}_2(\vec{r}_2,t_2)E^{(+)}_1(\vec{r}_1,t_1)\vec{\rho}_{12}],\label{S1}
\end{eqnarray}
\end{widetext}
where $E^{(\pm)}_1(\vec{r}_1,t_1)$ and $E^{(\pm)}_2(\vec{r}_2,t_2)$ are the quantized field operators at detector D1 and D2 located at space-time locations $(\vec{r}_1,t_1)$ and $(\vec{r}_2,t_2)$, respectively. For a quantum system which can be written in a pure state $|\Psi\rangle_{12}$, the second-order correlation function can be simplified to 
\begin{eqnarray}
G^{(2)}(\vec{r}_1,\vec{r}_2,t_1,t_2) &=& |\langle0|E^{(+)}_1(\vec{r}_1,t_1)E^{(+)}_2(\vec{r}_2,t_2)|\Psi\rangle_{12}|^2 \nonumber \\
 &=& |\mathcal{A}_{12}(\vec{r}_1,\vec{r}_2,t_1,t_2)|^2,\label{S2}
\end{eqnarray}
where $\mathcal{A}_{12}(\vec{r}_1,\vec{r}_2,t_1,t_2)$ is defined as the two-photon amplitude. 

\section{Einstein-Podolsky-Rosen position-momentum entanglement}

Let us consider a pure state that describes a system of photon pair entangled in transverse position and momentum variables. The state can be written in transverse momentum basis as \cite{D'Angelo2005},
\begin{equation}
|\Psi\rangle_{12}=\int{d\vec{\kappa}_1d\vec{\kappa}_2}\tilde{\mathcal{E}}_+\left(|\vec{\kappa}_1+\vec{\kappa}_2|\right)\tilde{\mathcal{E}}_-\left({|\vec{\kappa}_1-\vec{\kappa}_2|}/{2}\right)|\vec{\kappa}_1\vec{\kappa}_2\rangle,\label{S3}
\end{equation} 
where $\vec{\kappa}_j$ is transverse component of wave vector which is proportional to transverse momentum ($\vec{p}_j=\hbar\vec{\kappa}_j$) and $\tilde{\mathcal{E}}_{\pm}$ are assumed to be Gaussian functions defined as,
\begin{eqnarray}
&\phantom&\tilde{\mathcal{E}}_+(|\vec{\kappa}_{1}+\vec{\kappa}_{2}|)=\frac{1}{ 
\sqrt[\leftroot{-3}\uproot{3}4]{\pi\sigma_+^2}}\,\mathrm{exp}\left({-\frac{|\vec{\kappa}_{1}+\vec{\kappa}_{2}|^2}{2\sigma_+^2}}\right), \nonumber\\
&\phantom&\tilde{\mathcal{E}}_-(|\vec{\kappa}_{1}-\vec{\kappa}_{2}|/2)=\frac{1}{ 
\sqrt[\leftroot{-3}\uproot{3}4]{\pi\sigma_-^2}}\,\mathrm{exp}\left({-\frac{(|\vec{\kappa}_{1}-\vec{\kappa}_{2}|/2)^2}{2\sigma_-^2}}\right),~~~~~\label{envelope}
\end{eqnarray}
where $\sigma_{+}$ and $\sigma_{-}$ are the standard deviations of $\tilde{\mathcal{E}}_+$ and $\tilde{\mathcal{E}}_-$, respectively. For the case of $\sigma_{+}\ll\sigma_{-}$, the state is strongly anti-correlated in momentum. The state can also be written in transverse position basis as \cite{D'Angelo2005},
\begin{equation}
|\Psi\rangle_{12}=\int{d\vec{\rho}_1d\vec{\rho}_2}\mathcal{E}_+\left({|\vec{\rho}_1+\vec{\rho}_2|}/{2}\right)\mathcal{E}_-\left(|\vec{\rho}_1-\vec{\rho}_2|\right)|\vec{\rho}_1\vec{\rho}_2\rangle,\label{S4}
\end{equation}
where $\vec{\rho}_j$ is the transverse position and $\mathcal{E}_{\pm}$ are the Fourier transforms of $\tilde{\mathcal{E}}_{\pm}$. The two-photon probability distributions, $|\langle\vec{\kappa}_1\vec{\kappa}_2|\Psi\rangle_{12}|^2$ and $|\langle\vec{\rho}_1\vec{\rho}_2|\Psi\rangle_{12}|^2$, give rise to two-photon joint uncertainty relation \cite{D'Angelo2005},
\begin{equation}
\Delta(\vec{p}_1+\vec{p}_2)=\frac{\hbar\sigma_+}{\sqrt{2}},~~\Delta(\vec{\rho}_1-\vec{\rho}_2)=\frac{1}{\sqrt{2}\,\sigma_-},\label{S5}
\end{equation}
where the uncertainty of the total momentum is $$\Delta(\vec{p}_1+\vec{p}_2)=\sqrt{\left\langle(\vec{p}_1+\vec{p}_2)^2\right\rangle-\left(\langle(\vec{p}_1+\vec{p}_2)\rangle\right)^2}$$ and the uncertainty of the relative position is $$\Delta(\vec{\rho}_1-\vec{\rho}_2) = \sqrt{\left\langle(\vec{\rho}_1-\vec{\rho}_2)^2\right\rangle-\left(\langle(\vec{\rho}_1-\vec{\rho}_2)\rangle\right)^2}.$$ Here, $\langle x\rangle$ denotes the expectation value of the variable $x$ given the quantum state $|\Psi\rangle_{12}$.

Therefore, it is possible to obtain the uncertainties of momentum sum and position difference by measuring the standard deviations $\sigma_{\pm}$ of the envelopes $\tilde{\mathcal{E}}_{\pm}$. The two-photon joint uncertainty relation is directly related to the conditional probability of inferring position or momentum of photon 2 given the position or momentum of photon 1 in the original EPR context \cite{Einstein1935}.


\section{Quantum state of SFWM}

The quantum state generated from SFWM can be written as \cite{Du2008,Cho2014}
\begin{eqnarray}
|\Psi\rangle=A&\displaystyle{\int}&{d\omega_{as}}{d\vec{\kappa}_{s}d\vec{\kappa}_{as}}\chi^{(3)}(\omega_{as},\omega_{s}) \mathrm{sinc}(\Delta k L/2)\nonumber\\
&\phantom&\times\tilde{\mathcal{E}}_+(|\vec{\kappa}_s+\vec{\kappa}_{as}|)\hat{a}^{\dagger}_{\vec{\kappa}_s} \hat{a}^{\dagger}_{\vec{\kappa}_{as}}|0\rangle,\label{S6}
\end{eqnarray} 
where $A$ is a normalization constant, $\chi^{(3)}(\omega_{as},\omega_{s})$ is the third-order nonlinear susceptibility of the medium, $\Delta k=(\vec{k}_p+\vec{k}_c-\vec{k}_s-\vec{k}_{as})\cdot \vec{z}$ is the longitudinal phase mismatch along the direction $\vec z$ of the 2D MOT of length $L$, and $\tilde{\mathcal{E}}_+(|\vec{\kappa}_{as}+\vec{\kappa}_{s}|)$ is the Fourier transform of the Gaussian pump transverse profile. $\hat{a}^{\dagger}_{\vec{\kappa}_s}$ and $\hat{a}^{\dagger}_{\vec{\kappa}_{as}}$ are the creation operators of photons with transverse wave vectors $\vec{\kappa}_s$ and $\vec{\kappa}_{as}$, respectively. The energy conservation condition confirms that $\omega_s=\omega_p+\omega_c-\omega_{as}$. Here, it is assumed that the coupling field is a plane wave with direction $\vec{k}_c$ and the medium is larger than the spatial envelope of the pump. 

By comparing Eq. (\ref{S3}) and Eq. (\ref{S6}), it is easy to see that and $\tilde{\mathcal{E}}_-$ in Eq. (\ref{S3}) is assumed to be constant in Eq. (\ref{S6}). Therefore, the quantum state of SFWM in the form written in Eq. (\ref{S6}) assumes that the variance in $\vec{\kappa}_{s}-\vec{\kappa}_{as}$ is infinity, which is hardly realistic in experiments. Therefore, we can assume that, without losing the generality, a Gaussian envelope $\tilde{\mathcal{E}}_-(|\vec{\kappa}_{as}-\vec{\kappa}_{s}|/2)=\mathrm{exp}\left[{-(|\vec{\kappa}_s-\vec{\kappa}_{as}|/2)^2/2\sigma_-^2}\right]$ is multiplied to the quantum state, since a constant is a special case of $\tilde{\mathcal{E}}_-$, when $\sigma_-\rightarrow\infty$. The quantum state of SFWM can then be written as,
\begin{eqnarray}
|\Psi\rangle=A&\displaystyle{\int}&{d\omega_{as}}{d\vec{\kappa}_{s}d\vec{\kappa}_{as}}\chi^{(3)}(\omega_{as},\omega_{s}) \mathrm{sinc}(\Delta k L/2)\nonumber\\
&\phantom&\times\mathcal{C}_{\bot}(\vec{\kappa}_+,\vec{\kappa}_-)\hat{a}^{\dagger}_{\vec{\kappa}_s} \hat{a}^{\dagger}_{\vec{\kappa}_{as}}|0\rangle,\label{S7}
\end{eqnarray} 
where $\mathcal{C}_{\bot}(\vec{\kappa}_+,\vec{\kappa}_-)=\tilde{\mathcal{E}}_+(|\vec{\kappa}_+|) \tilde{\mathcal{E}}_-(|\vec{\kappa}_{-}|/2)$ and  $\vec{\kappa}_{\pm}=\vec{\kappa}_{as}\pm\vec{\kappa}_{s}$.

\section{Calculation for ghost interference}
\begin{figure}[h]
\centering
\includegraphics[width=3.4in]{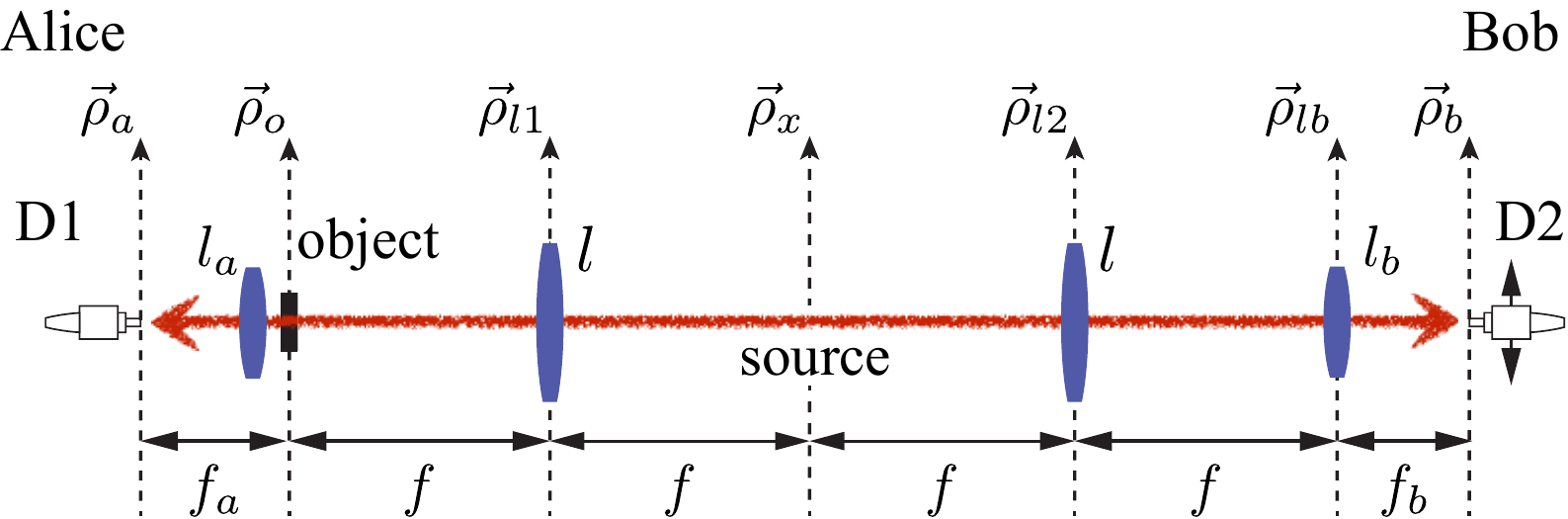}
\caption{The schematic of the ghost interference setup.}\label{interference}
\end{figure}

First, we begin by introducing an experimental set-up for ghost interference, as shown in Fig.~\ref{interference}. The narrowband photon pairs are generated from a cold atomic ensemble. One of the two-photon (Stokes) is sent to Alice and the other (Anti-Stokes) is sent to Bob, respectively. A pair of lenses ($l$) is installed in both Alice's and Bob's arms to optically relay the generated photons from the source to Alice's and Bob's measurement planes. Hence, the lenses enable efficient collection of generated photons from cold atomic ensemble. Also, the lenses are used to map either the position or the momentum of the photons at the output surface of the source to the detection plane. In Bob arm, the photons from the output surface of cold atom, $\vec{\rho}_x$, is traversing two lenses ($l$ and $l_b$) located at the distance $f$ and $2f$ from the source.
For the ghost interference, Bob's measurement plane is located at distance $f_b$ away from the lens $l_b$. Bob places a scanning single-mode fiber (SMF) mounted on a translation stage.
Here, the axes at the planes of lens $l$, $l_b$ and the plane of Bob's measurement are assigned to $\vec{\rho}_{l2}$, $\vec{\rho}_{lb}$, $\vec{\rho}_b$. 

On Alice's side, the photon transmits through the lens $l$ at $f$ away from the source. Right after an object, which is placed at the distance $2f$ from the source (or distance $f$ from the lens $l$), there is an objective lens $l_a$. Then Alice's measurement plane is placed at distance $f_a$ from the object. The axes of the planes of lens $l$, object, and Alice's measurement are assigned as $\vec{\rho}_{l1}$, $\vec{\rho}_{o}$, $\vec{\rho}_{a}$. Alice places a SMF at the center of the measurement plane $\vec{\rho}_{a}=0$. 

The quantized field operators at the detection planes can be represented in terms of the fields on the output planes of the source by using optical transfer function. In the paraxial approximation, the field can be written as \cite{D'Angelo2005,Rubin1996},
\begin{equation}
E^{(+)}_j(\vec{r}_j,t_j)=C\int{d\omega}{d\vec{\kappa}_j}\mathrm{exp}({-i\omega t_j}) g_j(\vec{\kappa}_j,\omega;\vec{\rho}_j,z_j)a_{\vec{\kappa}_j},\label{S8}
\end{equation}
where $\vec{r}=(\vec{\rho}_j,z_j)$ represents the position of detector $j$, $t_j$ is the time when the detector $j$ clicks, C is the normalization constant, $a_{\vec{k}}$ is the annihilation operator of a photon with momentum $\vec{k}=(\vec{\kappa},k_z)$, and $g_j(\vec{\kappa},\omega;\vec{\rho}_j,z_j)$ is the optical transfer function for the corresponding arm $j$ of the setup.

To calculate the field at Bob's detection plane $\vec{\rho}_b$, $E^{(+)}_b(\vec{\rho}_b,z_b,t_b)$, it is required to calculate the optical transfer function including two lenses, $l$ and $l_b$ with focal lengths $f$ and $f_b$, respectively. Here, $z_b=2f+f_b$, and the field at Bob's detection plane can be written as,
\begin{widetext}
\begin{equation}
E^{(+)}_b(\vec{\rho}_b,z_b,t_b)=C_b\displaystyle{\int}{d\omega}{d\vec{\kappa}_b}\mathrm{exp}({-i\omega(t_b-z_b/c)})G(|\vec{\rho}_b|)_{[\omega/cf_b]}\mathrm{exp}({-i\frac{f}{f_b}\vec{\kappa}_b\cdot\vec{\rho}_b})a_{\vec{\kappa}_b},
\label{S9}
\end{equation}
where $C_b$ is the normalization constant and $G(|\alpha|)_{[\beta]}$ is a Gaussian function defined as $G(|\alpha|)_{[\beta]}=\mathrm{exp}({i\beta/2|\alpha|^2})$.

Similarly, by calculating the optical transfer function on Alice's arm, we can calculate the field at Alice's detection plane $E^{(+)}_a(\vec{\rho}_a,z_a,t_a)$, where $z_a=2f+f_a$. The focal lengths of lens $l$ and $l_a$ are assumed to be $f$ and $f_a$, respectively. The object imposes a constraint in transverse space at plane $\vec{\rho}_o$, which can be represented as transmittance function $\mathcal{T}(\vec{\rho}_o)$. For $d''=d_o=f''=f$, the field at Alice's detection plane can be calculated as,
\begin{equation}
E^{(+)}_a(\vec{\rho}_a,z_a,t_a)=C_a\displaystyle{\int}{d\omega}{d\vec{\kappa}_a}\mathrm{exp}({-i\omega(t_a-z_a/c)})G(|\vec{\rho}_a|)_{[\omega/cf_a]}\mathcal{T}(\frac{\lambda f}{2\pi}\vec{\kappa}_a)\mathrm{exp}({-i\frac{f}{f_a}\vec{\kappa}_a\cdot\vec{\rho}_a})a_{\vec{\kappa}_a},
\label{S10}
\end{equation}
where $C_a$ is the normalization constant, $\lambda$ is the wavelength of the photon.

The two-photon amplitude can be calculated from Eq. (\ref{S7}), Eq. (\ref{S9}) and Eq. (\ref{S10}):
\begin{eqnarray}
\mathcal{A}_{ab}(\vec{r}_a,\vec{r}_b,t_a,t_b)&=&\langle0|E^{(+)}_a(\vec{r}_a,t_a)E^{(+)}_b(\vec{r}_b,t_b)|\Psi\rangle\nonumber\\
&=&C_1\int{d\omega_{as}}{d\vec{\kappa}_{s}d\vec{\kappa}_{as}}\chi^{(3)}(\omega_{as},\omega_{s}) \mathrm{sinc}(\Delta k L/2)\mathrm{exp}({-i\frac{f}{f_a}\vec{\kappa}_{s}\cdot\vec{\rho}_a})\nonumber\\
&\phantom&~~~~~~\times\mathrm{exp}({-i\frac{f}{f_b}\vec{\kappa}_{as}\cdot\vec{\rho}_b})\mathcal{C}_{\bot}(\vec{\kappa}_+,\vec{\kappa}_-)\mathcal{T}(\frac{\lambda_s f}{2\pi}\vec{\kappa}_{s}),\label{S11}
\end{eqnarray}
where $\vec{\kappa}_{\pm}=\vec{\kappa}_{s}\pm\vec{\kappa}_{as}$, $\lambda_s$ is the wavelength of the Stokes photon, and $C_1$ is a constant including all time-varying phase terms and irrelevant Gaussian functions, $G(|\vec{\rho}_a|)_{[\omega/cf_a]}$ and $G(|\vec{\rho}_b|)_{[\omega/cf_b]}$. We can calculate the two-photon correlation function by inserting Eq. (\ref{S11}) into Eq. (\ref{S2}). The transverse component of the two-photon correlation function can be written as,
\begin{eqnarray}
G^{(2)}(\vec{r}_a,\vec{r}_b,t_a,t_b)&=&|\mathcal{A}_{ab}(\vec{r}_a,\vec{r}_b,t_a,t_b)|^2 \nonumber\\
&\propto&\left|\int{d\vec{\kappa}_{s}d\vec{\kappa}_{as}}\mathrm{exp}({-i\frac{f}{f_a}\vec{\kappa}_{s}\cdot\vec{\rho}_a})\mathrm{exp}({-i\frac{f}{f_b}\vec{\kappa}_{as}\cdot\vec{\rho}_b})
\mathcal{C}_{\bot}(\vec{\kappa}_+,\vec{\kappa}_-)\mathcal{T}(\frac{\lambda_s f}{2\pi}\vec{\kappa}_{s})\right|^2.\label{S12}
\end{eqnarray}
When the two-photon state has ideal EPR entanglement, in other words $\mathcal{C}_{\bot}(\vec{\kappa}_+,\vec{\kappa}_-)=\delta(|\vec{\kappa}_{as}+\vec{\kappa}_{s}|)$, a simple calculation result in,
\begin{eqnarray}
G^{(2)}(\vec{r}_a,\vec{r}_b,t_a,t_b)\propto\left|\tilde{\mathcal{T}}(\frac{f}{f_a}\vec{\rho}_a+\frac{f}{f_b}\vec{\rho}_b)\right|^2,\label{S13}
\end{eqnarray}
\end{widetext}
which is simply proportional to Fourier transform of the object transfer function. For $\vec{\rho}_a=0$, the Fourier transform of the object transfer function is reconstructed in Bob's plane, $|\tilde{\mathcal{T}}(\frac{f}{f_b}\vec{\rho}_b)|^2$, which exhibits ghost interference. 

In a more realistic situation considering the finite pump spatial envelope and the divergence of the generated photons, $\mathcal{C}_{\bot}(\vec{\kappa}_+,\vec{\kappa}_-)$ becomes inequivalent to the delta function. The two-photon correlation function thus deviates from the ideal Fourier transform. The integration in Eq. (\ref{S12}) can in general be calculated numerically, but for a specific form of the object transfer function ${\mathcal{T}}(\vec{\rho}_o)$, an analytic expression of the integration can be calculated. In our case, the effective object transfer function is the product of the object (vertically aligned metal block of width $w_b$) and the Gaussian function defined by the objective lens ($l_a$) and the SMF. In the horizontal axis of the object plane, the effective object transfer function is ${\mathcal{T}}({\rho}_o)=c_o\mathrm{exp}(-\rho_o^2/w_0^2)(1-\Pi(\rho_o/w_b))$, where $w_0$ is the Gaussian envelope, $\Pi(x)=H(x+1/2)-H(x-1/2)$, $H(x)$ is the Heaviside step function. One can calculate the analytic integration of Eq. (\ref{S12}) using the effective double-slit object transfer function ${\mathcal{T}}({\rho}_o)$. For simplicity of the calculation, let us assume that the wavelength of the Stokes and Anti-Stokes are degenerate. Assuming that the Alice's detector is located at $\rho_a=0$, the calculation result in,
\begin{widetext}
\begin{eqnarray}
G^{(2)}(\rho_b)&\propto& \left|\frac{\sigma_+\sigma_-w_0}{\sqrt{8\pi^2w_0^2+f^2(\sigma_+^2+4\sigma_-^2)\lambda^2}}\mathrm{exp}\left[-\frac{f^2\rho_b^2(\pi^2(\sigma_+^2+4\sigma_-^2)w_0^2+2f^2\sigma_+^2\sigma_-^2\lambda^2)}{f_b^2(8\pi^2w_0^2+f^2(\sigma_+^2+4\sigma_-^2)\lambda^2)}\right]\right. \nonumber\\
&\phantom&\times\left(\mathrm{erfc}\left[\frac{-2if^2\pi (\sigma_+^2-4\sigma_-^2)w_0^2\lambda\rho_b+f_b w_b(8\pi^2 w_0^2+f^2(\sigma_+^2+4\sigma_-^2)\lambda^2)}{2f f_b w_0 \lambda\sqrt{(\sigma_+^2+4\sigma_-^2)(8\pi^2 w_0^2+ f^2(\sigma_+^2+4\sigma_-^2)\lambda^2)}}\right]\right.\nonumber\\
&\phantom&~~~~~~+\left.\left.\mathrm{erfc}\left[\frac{2if^2\pi (\sigma_+^2-4\sigma_-^2)w_0^2\lambda\rho_b+f_b w_b(8\pi^2 w_0^2+f^2(\sigma_+^2+4\sigma_-^2)\lambda^2)}{2f f_b w_0 \lambda\sqrt{(\sigma_+^2+4\sigma_-^2)(8\pi^2 w_0^2+ f^2(\sigma_+^2+4\sigma_-^2)\lambda^2)}}\right]\right)\right|^2,~~~~~~~~\label{S14}
\end{eqnarray}
\end{widetext}
where $f$, $f_b$ are the focal lengths of the lenses $l$, $l_b$, $\lambda$ is the wavelength of the photon, $\mathrm{erfc}(x)$ is the complementary error function defined as $\mathrm{erfc}(x)=2(\sqrt{\pi})^{-1}\int_x^{\infty}\mathrm{exp}(-t^2)dt$. By fitting the experimental data to Eq. (\ref{S14}), it is possible to obtain $\sigma_{+}$ and $\sigma_{-}$. Following Eq. (\ref{S5}), the standard deviation of the total momentum, $\Delta p_+$, can be calculated as $\Delta p_+=\hbar \sigma_{+}/\sqrt{2}$. Similarly, the standard deviation of the transverse position, $\Delta x_-$, can be calculated as $\Delta x_-=\sigma_{-}^{-1}/\sqrt{2}$.

In the ghost interference measurement, the fitting parameters $\sigma_{+}$ and $\sigma_{-}$ are obtained by assuming a pure state notation of the quantum state as in Eq. (\ref{S7}). Note that the high visibility ghost interference \textit{alone} may not be a conclusive evidence of EPR entanglement, because it is possible to simulate either ghost interference or ghost imaging with a classically correlated, separable light source \cite{Bennink2002,Bennink2004,Gatti2004,D'Angelo2005}. However, it is fundamentally impossible to simulate \textit{both} ghost interference \textit{and} ghost imaging with classically correlated light source \cite{Bennink2004,D'Angelo2005}. Therefore, it is necessary to observe the high contrast ghost imaging as well as the high visibility ghost interference in order to conclusively confirm EPR entanglement.


\section{Calculation for ghost imaging}

\begin{figure}[b]
\centering
\includegraphics[width=3.4in]{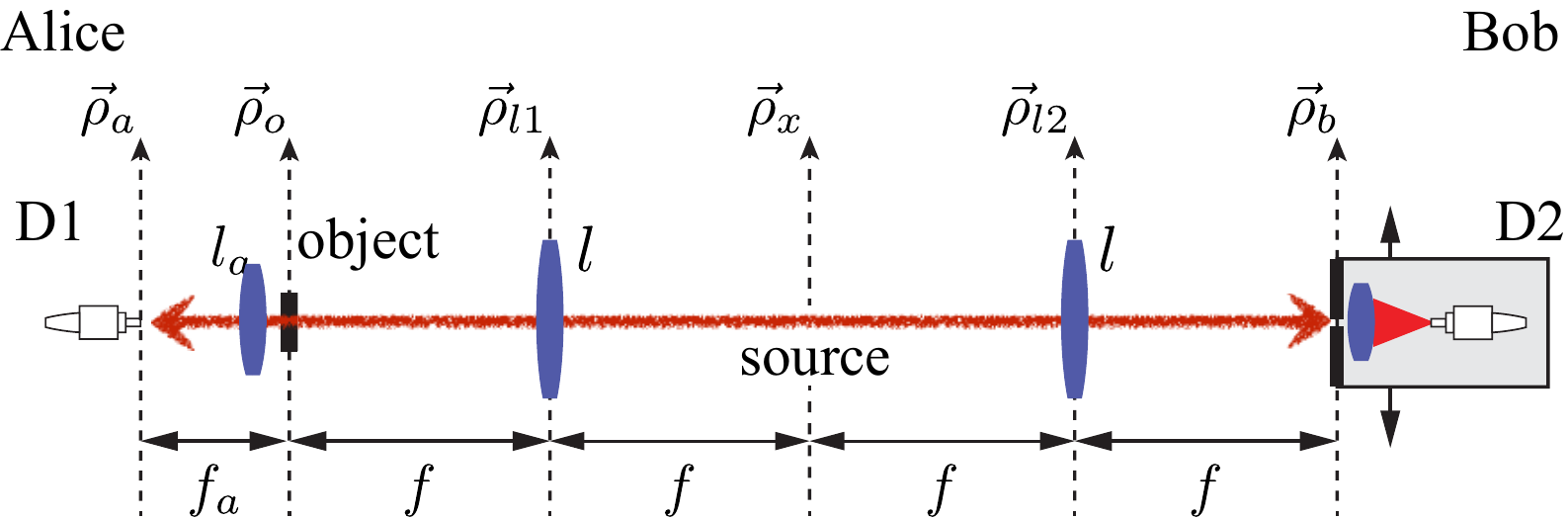}
\caption{The schematic of the ghost imaging setup.}\label{imaging}
\end{figure}

Now we consider the two-photon correlation function for ghost imaging experimental set-up, as shown in Fig. \ref{imaging}. Here, Alice's arm is the same as that of ghost interference set-up; hence, the quantized field operator at Alice's detection plane is the same as Eq. (\ref{S10}). On the other hand, the Bob's arm has been modified from ghost interference to scan the far-field or Fourier plane of the source.

\begin{widetext}
The quantized field operator at Bob's plane $\vec{\rho}_b$ can be calculated as,
\begin{eqnarray}
E^{(+)}_b(\vec{\rho}_b,z_b,t_b)=C_b\displaystyle{\int}{d\omega}{d\vec{\kappa}_b}\mathrm{exp}({-i\omega(t_b-z_b/c)})G(|\vec{\rho}_b|)_{[\omega/cf]}
a_{\vec{\kappa}_b}G(|\vec{\kappa}_b|)_{[-cf/\omega]}\delta(\vec{\kappa}_b-\frac{\omega}{cf}\vec{\rho}_b),
\label{S15}
\end{eqnarray}
where $C_b$ is a normalization constant and $z_b=2f$. 

The two-photon amplitude can be calculated by using Eq. (\ref{S7}), (\ref{S10}) and (\ref{S15}):
\begin{eqnarray}
\mathcal{A}_{ab}(\vec{r}_a,\vec{r}_b,t_a,t_b)&=&C_2\int{d\omega_{as}}{d\vec{\kappa}_{s}d\vec{\kappa}_{as}}\chi^{(3)}(\omega_{as},\omega_s) \mathrm{sinc}(\Delta k L/2)\nonumber\\
&\phantom&\times \mathrm{exp}({-i\frac{f}{f_a}\vec{\kappa}_{s}\cdot\vec{\rho}_a})
\mathcal{C}_{\bot}(\vec{\kappa}_+,\vec{\kappa}_-)\mathcal{T}(\frac{\lambda_s f}{2\pi}\vec{\kappa}_{s})G(|\vec{\kappa}_{as}|)_{[-cf/\omega]}\delta(\vec{\kappa}_{as}-\frac{\omega}{cf}\vec{\rho}_{b}),~~~~~~~\label{S16}
\end{eqnarray}
where $C_2$ is a constant including all time-varying phase terms and irrelevant Gaussian functions, $G(|\vec{\rho}_a|)_{[\omega/cf_a]}$ and $G(|\vec{\rho}_b|)_{[\omega/cf]}$.  The transverse component of two-photon correlation function for ghost imaging can be calculated to be,
\begin{eqnarray}
&\phantom&G^{(2)}(\vec{r}_a,\vec{r}_b,t_a,t_b)=|\mathcal{A}_{ab}(\vec{r}_a,\vec{r}_b,t_a,t_b)|^2 \nonumber\\
&\phantom&~~\propto\left|\int{d\vec{\kappa}_{s}d\vec{\kappa}_{as}}\mathrm{exp}({-i\frac{f}{f_a}\vec{\kappa}_{s}\cdot\vec{\rho}_a})
\mathcal{C}_{\bot}(\vec{\kappa}_+,\vec{\kappa}_-)\mathcal{T}(\frac{\lambda_s f}{2\pi}\vec{\kappa}_{s})G(|\vec{\kappa}_{as}|)_{[-cf/\omega]}\delta(\vec{\kappa}_{as}-\frac{\omega}{cf}\vec{\rho}_{b})\right|^2.~~~~~\label{S17}
\end{eqnarray}
\end{widetext}
Again, when the two-photon state has ideal EPR entanglement, i.e. $\mathcal{C}_{\bot}(\vec{\kappa}_+,\vec{\kappa}_-)=\delta(|\vec{\kappa}_{as}+\vec{\kappa}_{s}|)$, the two-photon correlation is reduced to,
\begin{eqnarray}
G^{(2)}(\vec{r}_a,\vec{r}_b,t_a,t_b)\propto\left|\mathcal{T}(-\vec{\rho}_b)\right|^2,\label{S18}
\end{eqnarray}
which is proportional to the object transfer function itself reconstructed in Bob's plane ($\vec{\rho}_b$).

When the EPR entanglement is non-ideal, i.e. $\mathcal{C}_{\bot}(\vec{\kappa}_+,\vec{\kappa}_-)$ is not a delta function, the two-photon correlation function has to be calculated from Eq. (\ref{S17}), which likely give blurred image of the object. Similarly to the ghost interference case, by assuming the effective object transfer function ${\mathcal{T}}({\rho}_o)=c_o\mathrm{exp}(-\rho_o^2/w_0^2)(1-\Pi(\rho_o/w_b))$, one can analytically integrate Eq. (\ref{S17}). For simplicity of the calculation, let us assume that the wavelength of the Stokes and Anti-Stokes are degenerate. Assuming that the Alice's detector is located at $\rho_a=0$, the calculation result in,
\begin{widetext}
\begin{eqnarray}
G^{(2)}(\rho_b)&\propto& \left|\left.\frac{\sigma_+\sigma_-w_0}{\sqrt{2 \pi^2 (\sigma_+^2+4\sigma_-^2)w_0^2+4 f^2 \sigma_+^2 \sigma_-^2 \lambda^2}}\mathrm{exp}\left[-\frac{2 \pi^2 \rho_b^2(8\pi^2 w_0^2+f^2(\sigma_+^2+4\sigma_-^2)\lambda^2)}{f^2\lambda^2(2 \pi^2 (\sigma_+^2+4\sigma_-^2) w_0^2 + 4 f^2 \sigma_+^2\sigma_-^2\lambda^2)}\right]\right.\right. \nonumber\\
&\phantom&~\times\left(2-\mathrm{erf}\left[\frac{2 f_a \pi^2 w_0^2\left(4\sigma_-^2(w_b-2\rho_b)+\sigma_+^2(w_b+2\rho_b)\right)+4 f^2 f_a \sigma_+^2 \sigma_-^2 w_b \lambda^2}{4f f_a \sigma_+\sigma_-w_0\lambda\sqrt{2\pi^2(\sigma_+^2+4\sigma_-^2)w_0^2+4 f^2 \sigma_+^2 \sigma_-^2 \lambda^2}}\right]\right.\nonumber\\
&\phantom&~~~~~~~\left.\left.- \mathrm{erf}\left[\frac{2 f_a \pi^2 w_0^2\left(4\sigma_-^2(w_b+2\rho_b)+\sigma_+^2(w_b-2\rho_b)\right)+4 f^2 f_a \sigma_+^2 \sigma_-^2 w_b \lambda^2}{4f f_a \sigma_+\sigma_-w_0\lambda\sqrt{2\pi^2(\sigma_+^2+4\sigma_-^2)w_0^2+4 f^2 \sigma_+^2 \sigma_-^2 \lambda^2}}\right]\right)\right|^2.~~~~~~~~\label{S19}
\end{eqnarray}
\end{widetext}
By fitting the ghost imaging experimental data to Eq. (\ref{S19}), one can obtain $\sigma_{+}$ and $\sigma_{-}$ and calculate the joint uncertainties $\Delta p_+$ and $\Delta x_-$ using Eq. (\ref{S5}).